# Two-dimensional elasticity determines the low-frequency dynamics of single- and double-walled carbon nanotubes

S.B. Rochal[1,2], V.L. Lorman[2], Yu.I. Yuzyuk[1]

[1]Faculty of Physics, Southern Federal University, 5 Zorge str., 344090 Rostov-on-Don, Russia
[2]Laboratoire Charles Coulomb, UMR 5221 CNRS – Université Montpellier 2, pl. E. Bataillon, 34095 Montpellier, France

We develop a continuous theory of low-frequency dynamics for nanotubes with truly two-dimensional (2D) walls constituted by single-atom monolayer. In this frame topological bending elasticity of the monolayer is not related to its vanishing macroscopic thickness. The proposed approach predicts completely new sound dispersions and radius dependences of non-resonant Raman-active modes frequencies in single-walled carbon nanotubes (SWCNT). Resulting relations are suitable for nanotubes identification and more complete or alternative characterization. The theory is also applied to describe the low-frequency dynamics of double-walled carbon nanotubes (DWCNT). It establishes a clear-cut relation between the radial breathing mode in SWCNT and breathing-like modes in DWCNT and fits the existing Raman data better than previously developed 3D continuous or discrete models. The results obtained constitute the basis for new quantitative studies of the low-frequency vibrational spectrum, heat capacity and heat transfer properties of carbon nanotubes.

Vibrational properties of carbon nanotubes are strongly dependent on phonon confinement and cylindrical geometry. Their measurements provide a powerful tool to get an insight into crucial nanotube characteristics. The low-frequency region of the vibrational spectrum is the most important part for sample identification and characterization of its dimensional parameters. Indeed, the frequency of radial breathing mode (RBM) of a freely suspended single-walled carbon nanotube (SWCNT) is with a good accuracy proportional to its inverse diameter. The relation is practically independent on SWCNT type, i.e. on the way the SWCNT net is «cut» in a graphene sheet. In addition, radial vibrations like RBM involve homogeneous tube strain only. These facts have served as starting points for more general ideas suggesting that the continuous theory of thin cylindrical shell dynamics can be suitable to study vibrational modes (i.e. to calculate frequencies and visualize displacement fields) of nanotubes involving homogeneous or weakly inhomogeneous strain.  The continuous theory of SWCNT dynamics proposed about a decade ago has started with exactly these ideas [1,2]. The theory does not deal with the high-frequency region of the vibrational spectrum. To study high-frequency optic-like modes relevant models take into account discrete atomic structure of the nanotube (see, for example [3]). In spite of limitations of its applicability region, the continuous theory is strongly attractive due to several evident advantages. In this frame the equations of motion can be solved analytically, and the resulting rather simple expressions for mode frequencies can be easily used for low-frequency Raman spectroscopy data fitting.

Recent encouraging progress in the Raman spectroscopy technique [4] indicates that in near future it will become possible to measure Raman-active but not resonant modes of SWCNT. Continuous theory results will constitute the basis for fitting of these qualitatively new data, and consequently, for more complete or alternative methods for nanotube characterization. Another domain where the advantages of continuous theory lead to an evident progress is Raman studies of breathing-like modes (BLM) in double-walled carbon nanotubes (DWCNT) [5] and in triple-walled nanostructures [6], already actively investigated.

Continuous theory of SWCNT proposed in [1,2] managed to relate longitudinal acoustic wave velocity in a thin graphite plate with the RBM frequency in SWCNT. However, the predictions of this theory for other low-frequency vibrational modes seem to be not completely justified. The problem is related to the fact that graphene sheet in the theory is considered as a finite-thickness plate, though thickness in the continuous theory of membrane dynamics is an essentially *macroscopic* parameter. Let us recall that continuous mechanics considers bending (or flexural) elasticity of a finite-thickness membrane as a result of the strain difference in the membrane's parts situated above and below its mid-surface. This relation is no more applicable to continuous mechanics of a membrane constituted by a single-atom layer. In a self-consistent continuous model single-layer graphene sheet should be considered as a 2D membrane. Corresponding bending elasticity of the 2D membrane is a purely topological quantity related to its curvature variation. Along the same line, all characteristics of the continuous mechanics theory for a 2D material membrane in a three-dimensional (3D) space are easily obtained using methods of differential geometry [7], in a way similar to classical 3D theory of elasticity [8].

Atoms (considered as material points) in the 2D surface located in the 3D space can move in three orthogonal directions. Thus, three low-amplitude waves can propagate simultaneously on the surface, namely, shear, stretching and bending waves. In the general case these modes are coupled. Three coupled equations of motion for corresponding modes in a SWCNT with a finite wall thickness were proposed in [2]. However, these results are not convincing enough. It is easy to see that in the equations obtained in [2] the Goldstone mode, which corresponds to the translation of the nanotube as a whole, has non-zero frequency. The same remark concerns another version of the SWCNT equations of motion applied to the low-temperature phonon heat capacity calculations [9]. In the Mahan's pioneering work [1] the membrane equations of motion were replaced by the general equations for an isotropic continuous 3D medium which were solved in cylindrical coordinates with the boundary conditions corresponding to a thin (but finite-thickness) cylindrical shell cut in this medium. Each series of modes was obtained by using additional approximations which are not the same for all mode types. Unique secular equation describing entire SWCNT low-frequency spectrum was not proposed and Raman activity of the modes was not analyzed correctly. Thus, elaboration of a self-consistent continuous 2D theory with unified conclusions about all modes present in this spectral region seems to be opportune as well as its application to Raman spectroscopy data analysis.

The aim of this Letter is double. First, we develop a 2D continuous theory of SWCNT dynamics for tubules with truly monoatomic walls, suitable for self-consistent description of non-resonant Raman-active modes in SWCNT. Second, we generalize the theory to the DWCNT case and apply for the interpretation of recent experimental data on BLM behavior.

To avoid evident difficulties of the limit transition from the 3D mechanical model of a thin finite-thickness plate to the model of a 2D monoatomic membrane we introduce from the very beginning the free energy density for a strained 2D material membrane in the 3D space based on the classical principles of elastic energy derivation [8]. Similar to the 3D case the approach involves topological and differential geometry properties of the medium. Minimal model for the 2D membrane strain energy density is then given by:

$$g = \frac{\lambda}{2}(\varepsilon_{ii})^2 + \mu \varepsilon_{ij}^2 + K(\Delta H)^2 \qquad (1)$$

where $\lambda$ and $\mu$ are 2D analogues of Lame coefficients, $K$ is topological bending rigidity, $\Delta H = H - H_0$, with $H_0$ standing for spontaneous mean curvature of the surface and $H$ being mean curvature resulting from the membrane strain, $\varepsilon_{ij}$ is 2D strain tensor with the components depending on the 3D displacement field **u** and $H_0$. Note that for 2D material membranes bending modulus $K$ is independent on the elastic constants $\lambda$ and $\mu$ characterizing in-plane strain components. This is in contrast with the 3D continuous theory which relates bending elasticity of a plate to its thickness, Young modulus and Poisson ratio. Equations of motion for the 2D material membrane are obtained by variation of functional:

$$A = \int \left( g(\mathbf{u}) - \rho \dot{\mathbf{u}}^2/2 \right) dS\, dt, \qquad (2)$$

where $t$ is time, $dS$ is the membrane area element, and $\rho$ is the surface mass density of the membrane. Solutions of the motion equations determine **u** as a function of space and time variables. Free energy density can be written in its general form (1) for both plane and cylindrical membranes, but the expressions for the 2D strain tensor $\varepsilon_{ij}$ and curvature deviation $\Delta H = H - H_0$ are quite different for these two geometries. For the plane membrane $H_0=0$, $H = \partial_x^2 u_z + \partial_y^2 u_z$, $\varepsilon_{xx} = \partial_x u_x$, $\varepsilon_{yy} = \partial_y u_y$ and $\varepsilon_{xy} = (\partial_x u_y + \partial_y u_x)/2$. Solutions of the corresponding equations of motion are superpositions of three independent waves. Two of them, iLA and iTA, do not break the plane symmetry of the membrane and are characterized by the velocities

$$V_{iLA} = \sqrt{(\lambda + 2\mu)/\rho}, \quad V_{iTA} = \sqrt{\mu/\rho}. \qquad (3)$$

The velocity of the third bending wave oLA is equal to zero, the mode being characterized by the quadratic dispersion law

$$\omega_{oTA} = k^2 \sqrt{K/\rho}, \qquad (4)$$

where $\omega_{oTA}$ - is the angular frequency of this wave and $k$ is the wave vector's length. Derivation of Eqs (3-4) is straightforward and can be easily obtained by simple generalization of the classical elasticity theory formulas [8].

Here we will focus on the dynamics of the 2D cylindrical membrane. Strain of such a membrane is characterized by a displacement field $\mathbf{u}=(u_r, u_\varphi, u_z)$, with 3 components parameterized in cylindrical coordinates by the angle $\varphi$ and the variable $l$, which measures the distance along the axis of the unstrained membrane. Symmetric 2D tensor of local strain $\varepsilon_{ij}$ is obtained using linearized deviation (induced by strain of a cylindrical surface with radius $R$) of the metric tensor [7] (see Appendix A) or by reducing classical 3D tensor [8] to the 2D case.

To derive the equations of motion of the cylindrical membrane we obtain curvature deviation $\Delta H$ linearized with respect to the displacement field components and their derivatives $\Delta H = -(\Delta_s u_r)/(2R^2)$, where $\Delta_s = 1 + \partial_\varphi^2 + R^2 \partial_l^2$, substitute $\Delta H$, cylinder area element $dS = Rdld\varphi$ and strain tensor components in (2) and calculate the variation. Resulting equations have the following form:

$$\ddot{u}_r \rho R = -(\lambda + 2\mu)\left(\frac{u_r}{R} + \frac{\partial u_\varphi}{R\partial \varphi}\right) - \lambda \frac{\partial u_z}{\partial l} - K \frac{1}{R^3} \Delta_s^2 u_r$$

$$\ddot{u}_\varphi \rho R = \frac{(\lambda + 2\mu)}{R}\left(\frac{\partial u_r}{\partial \varphi} + \frac{\partial^2 u_\varphi}{\partial \varphi^2}\right) + (\lambda + \mu)\frac{\partial^2 u_z}{\partial \varphi \partial l} + \mu R \frac{\partial^2 u_\varphi}{\partial l^2} \qquad (5)$$

$$\ddot{u}_z \rho R = (\lambda + \mu)\frac{\partial^2 u_\varphi}{\partial \varphi \partial l} + (\lambda + 2\mu)R\frac{\partial^2 u_z}{\partial l^2} + \lambda \frac{\partial u_r}{\partial l} + \mu \frac{\partial^2 u_z}{R \partial \varphi^2}$$

Solutions of system (5) are easy to obtain in contrast to cumbersome expressions [1] of the 3D continuous theory for finite-thickness cylindrical shells. The methods are similar to those of the classical lattice dynamics commonly used in condensed matter physics. Namely, by substituting $u_j = u_j^0 \exp(i(kl + n\varphi - \omega t))$ where *n* is integer wave number, *k* is 1D wave vector and $j = r, \varphi, z$, we obtain dynamic matrix of the system. Vanishing of its determinant

$$\begin{bmatrix} \frac{\lambda + 2\mu}{R} + \frac{KX^2}{R^3} - R\rho\omega^2 & i\frac{(\lambda + 2\mu)n}{R} & ik\lambda \\ -i\frac{(\lambda + 2\mu)n}{R} & \frac{(\lambda + 2\mu)n^2}{R} + \mu k^2 R - \rho\omega^2 R & (\lambda + \mu)nk \\ -ik\lambda & (\lambda + \mu)nk & (\lambda + 2\mu)k^2 R + \frac{\mu n^2}{R} - R\rho\omega^2 \end{bmatrix} \qquad (6)$$

where $X = (R^2 k^2 + n^2 - 1)$, determines three real dispersion laws $\omega_j = \omega_j(k,n)$. Imaginary values of nondiagonal blocks in (6) express π/2 phase shifts of radial component with respect to the tangent ones.

To apply the proposed theory to the low-frequency spectrum of SWCNT and to calculate frequencies of its vibrational modes from (6) we perform following estimations of graphene and nanotube material constants. Following [10] we take in-plane sound wave velocities in graphene to be $V_{iLA} = 21.3$ km/s and $V_{iTA} \approx 13.6$ km/s. Similar estimations were obtained in [11] together with the estimation of graphene sheet bending rigidity as K≈2.1 Ev. Taking in addition surface density $\rho \approx 0.762$ mg/m$^2$ we obtained the following values for reduced constants in the spectroscopic unit system $\lambda/\rho \approx 2400$ cm$^{-2}$nm$^2$, $\mu/\rho \approx 5200$ cm$^2$nm$^2$, $K/\rho \approx 12.5$ cm$^{-2}$nm$^4$. To calculate in the chosen unit system mode frequencies determined by (6), nanotube radius is taken in *nm* and the frequencies are obtained in *cm$^{-1}$*.

The most low-frequency modes (with non-zero wave vector) of a SWCNT correspond to the tube's rigid-body motions with amplitudes weakly inhomogeneous along its axis. In the limit of vanishing wave vector (*k->0*) three SWCNT vibrational modes considered in this paragraph become tube's Goldstone modes: rotation around its axis, translation as a whole along the tube axis, and translation as a whole in the direction normal to the axis. Two SWCNT dispersion branches with *n=0* and transverse rotational or longitudinal translational displacements are usual acoustic vibrations, their frequencies vanish linearly for vanishing wave vector k. Similar to the conclusion of the Mahan's model [1], SWCNT rotational mode velocity in the 2D membrane theory is equal to $V_{iTA}$ mode velocity (3) in graphene. All other results obtained in the present approach are quite different with respect to the 3D continuous theory conclusions [1,2,9]. For the longitudinal mode the velocity $V_{LA}$ is estimated as

$$\left.\frac{\partial \omega}{\partial k}\right|_{k=0} \approx \left(\frac{4\mu(\lambda+\mu)}{\rho(\lambda+2\mu)}\right)^{1/2} \approx 20,9 \ km/s \qquad (7)$$

To obtain expression (7) we assume
$$\lambda R^2 >> K \ \text{and} \ \mu R^2 >> K \qquad (8)$$
which is a very good approximation for SWCNT. Expression for longitudinal mode velocity (7) gets in the frame of the 2D membrane theory a clear-cut physical meaning. Namely, it coincides with the expression for the longitudinal sound wave velocity in a narrow strip cut in the graphene sheet. The third branch of the low-frequency SWCNT spectrum corresponds to bending vibrations with $n=1$. Its dispersion law is quadratic $\omega = \alpha k^2$. Using again the fact that bending rigidity in SWCNT is rather weak, i.e. taking approximation (8), we obtain $\alpha \approx V_{LA}R/\sqrt{2}$.

In the following consideration we will discuss in detail frequencies of optically-active modes for which $k=0$ and matrix (6) becomes quasi-diagonal. Shear modes of this type have dispersion laws $\omega_{shear}(n) = n\sqrt{\mu/\rho}/R \approx 144 n/d$ and are not coupled with other modes. Bending and stretching modes with the same $n$ value are linearly coupled and the type of the mode is determined only quantitatively by the ratio of $u_r^0/u_\varphi^0$ amplitudes. For $u_r^0 > u_\varphi^0$ the mode can be considered to be bending one. In the opposite case it is a stretching mode. Frequencies of this couple of modes are obtained from the equation $det \ M_2 = 0$, where $M_2$ is the upper left 2x2 minor of matrix (6). As it is easy to see, in particular cases of $n=0$ and $n=1$ that one of the solutions of the equation $det \ M_2 = 0$ vanishes, and corresponding modes become tube's Goldstone rotation (for $n=0$) or translation (for $n=1$) as whole.

Let us now show the correspondence between the modes of the 2D continuous membrane model and the modes of SWCNT with discrete physical symmetry. We will be interested mainly in Raman-active modes of SWCNT which can, in principle, be measured experimentally in the non-resonant scattering experiments. All modes with $n>0$ span two-dimensional $E_{ng}$ or $E_{nu}$ representations of the $D_{\infty h}$ symmetry group of the cylindrical membrane. SWCNT has discrete symmetry $D_{Nh}$ or $D_N$ with even $N$ [12]. All $N/2-1$ two-dimensional representations $E_{ng}$ and $E_{nu}$ of the $D_{\infty h}$ group (with $n=1,2... N/2-1$) are reduced in the SWCNT discrete symmetry group to two-dimensional representations with the same $n$ values. As it is well known Raman-active modes in nanotubes are $E_{1g}$ и $E_{2g}$ ones ($E_1$ и $E_2$ modes in chiral nanotubes) [12]. Thus, the modes of the 2D continuous membrane model with $n>2$ are reduced in SWCNT to the modes which cannot be Raman-active. The modes of continuous cylindrical membrane with $k=0$ и $n\leq 2$ are quite similar to those of SWCNT since the wave lengths in this case are sufficiently long with respect to the distance between neighboring carbon atoms. Consequently, the 2D continuous theory predicts frequencies of the Raman-active modes with a good accuracy.

Vanishing of the (1,1) element of matrix (6) determines the frequency of the RBM mode ($n=0$). Using approximation (8), which introduces very small error (less than 0.5 cm$^{-1}$), we get $\omega_{RBM} \approx V_{iLA}/R \approx 226/d$. This result is in accord with recent experimental data approximating RBM frequency dependence on the tube diameter as $\omega_{RBM} \approx 228/d$ [13]. Similar expression for the RBM frequency was obtained in previous works in the frame of 3D continuous theory [1,2]. In contrast to these works the 2D continuous membrane approach developed above relates the RBM frequency to the velocity $V_{iLA}$ in a graphene sheet, and not to the velocity of longitudinal sound wave in a thin (but finite-thickness) graphite plate. Though these two velocities have rather close numerical values their expressions in terms of the elastic constants are quite different. Velocity $V_{iLA}$ in the graphene sheet simply coincides with the velocity of the sound wave (with the corresponding polarization) in a bulk graphite sample.

Using approximation (8) we easily obtain from matrix (6) the expressions for other Raman-active modes of SWCNT. Frequencies of "stretching" modes (remember that stretching and bending modes are linearly coupled) with $n=1$ and $n=2$ have simple relations with the RBM mode frequency, $\omega_{strt}(1) \approx \sqrt{2}\omega_{RBM}$ and $\omega_{str}(2) \approx \sqrt{5}\omega_{RBM}$, respectively. The mode with $n=1$ is characterized by the displacement field with equal amplitudes of tangent and radial displacement components. This is a necessary condition to preserve the tube's center-of-mass position during oscillations of this type (see Fig. 1). Bending mode with $n=2$ and frequency $\omega_{bend}(2) \approx \dfrac{6\sqrt{K}}{\sqrt{5}R^2} \approx 38/d^2$ is the softest mode in SWCNT. Furthermore, frequencies of all bending modes vanish for vanishing bending rigidity constant $K \to 0$. This implies that minimal model of the SWCNT dynamics should necessarily take into account bending rigidity term (as it is the case of density (1)); for $K=0$ the system is simply unstable.

Displacement fields for six SWCNT modes with $k=0$ and $n \leq 2$ are presented in Fig. 1. In chiral nanotubes all these modes are Raman-active. In achiral nanotubes one should first determine the mode type ($g$ or $u$). Maybe the simplest way to do it is to have a look at their displacement fields presented in Fig. 1. Detailed symmetry classification of low-frequency Raman-active SWCNT modes can be found in [14]. Frequencies of all Raman-active modes considered here in the frame of 2D continuous membrane theory tend to zero with increasing tube diameter according to simple laws, thus making non-resonant modes (Fig. 1, b-f) suitable for SWCNT identification and more complete characterization.

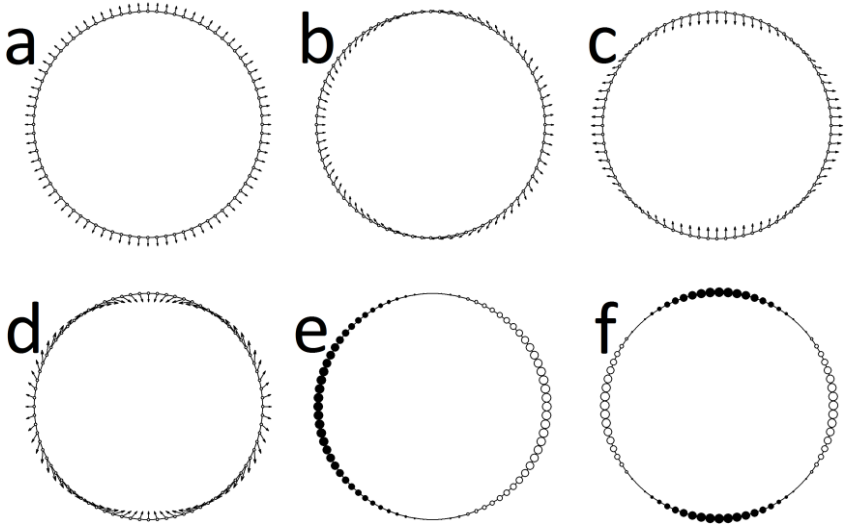

*Fig. 1 Displacement fields for low-frequency modes (with n=0,1,2) and zero wave vector. Modes (a-f) are the only Raman-active low-frequency modes in chiral SWCNT. In achiral nanotubes the only Raman-active low-frequency modes are limited to g modes (panels a, c, d, and e). Frequencies of the shown modes, including non-resonant modes (b-f), tend to zero with increasing SWCNT diameter according to simple laws.*
*(a): RBM; (b) stretching with n=1; (c-d) bending and stretching with n=2, respectively (note that the displacement field of the bending mode involves practically no membrane stretching); (e-f) shear modes with n=1 and n=2, corresponding displacement fields are perpendicular to the Fig. plane, displacements are presented by circles, displacement amplitude is given by the circle size, open or full circles correspond to up or down displacements, respectively.*

The 2D continuous membrane theory proposes also useful solutions which lead to progress in Raman studies of low-frequency modes in DWCNT. Generalizing the approach developed for SWCNT to the case of two co-axial weakly coupled SWCNT we obtain a clear description of the DWCNT low-frequency spectrum in terms of equations similar to Eqs. (5,6). To illustrate the results of the theory we analyze here BLM modes of DWCNT, actively discussed in recent experimental works [15-17]. The main contribution of van der Waals interaction of coupled co-axial SWCNT to the full free energy of the DWCNT system has the form:

$$U_{int} = \frac{G}{2} \int \left( u_r^{(1)} - u_r^{(2)} \right)^2 d\phi dl \qquad (9)$$

where $u_r^{(1)}$ and $u_r^{(2)}$ are radial displacement fields of the first and the second nanotubes, respectively. Similar terms dependent on tangent displacement fields can also be written, but their contributions to the energy of breathing-like modes are much smaller than (9) and can be omitted in the considered case. Full action of the system is then expressed as a sum of actions (2) of both nanotubes completed with coupling contribution (9): $A = A_1 + A_2 + \int U_{int} dt$. Following the procedure presented above in the case of SWCNT we obtain a system of six equations of motion describing low-frequency dynamics of DWCNT. Corresponding system is composed of two, slightly modified, copies of Eq. (5) for inner and outer tubes, respectively. Coupling term (9) gives the contribution $G(u_r^{(2)} - u_r^{(1)})$ to the right-hand side of the equation expressing radial field dynamics of the inner tube (first equation of the system), and $-G(u_r^{(2)} - u_r^{(1)})$ contribution to the equation of the outer tube radial dynamics (forth equation of the system).

Six-dimensional dynamic matrix *M* resulting from the equations of motion becomes quasi-diagonal and strongly simplified for certain simple modes. Secular equations obtained from the *det M = 0* condition (and using approximation (8) of weak bending rigidity, as we already did in the case of SWCNT) give simple dispersion laws and clear conclusions for the mode behavior. Namely, frequencies of two coupled BLM are found from the equation:

$$\begin{vmatrix} \frac{\lambda+2\mu}{R_1} - R_1\rho\omega^2 + G & -G \\ -G & \frac{\lambda+2\mu}{R_1} - R_1\rho\omega^2 + G \end{vmatrix} = 0 \qquad (10)$$

where $R_1$ and $R_2$ are the radii of the inner and outer tubes, respectively. Equation (10) constitutes a user-friendly tool for the BLM behavior analysis. For example, it is easy to compare BLM frequencies with the eigen frequencies of breathing modes of individual SWCNT (experimental comparison of these frequencies for a whole series of DWCNT is the subject of recent works [15,16]). For that aim we note eigen frequencies of the inner and outer tubes, as $\omega_1$ and $\omega_2$, respectively. Replacing $(\lambda + 2\mu)/R_i^2$ in Eq. (10) with $\omega_i^2$ and introducing $G' = G/\rho$ we obtain bi-quadratic equation

$$(R_1(\omega_1^2 - \omega^2) + G')(R_2(\omega_2^2 - \omega^2) + G') - G'^2 = 0, \qquad (11)$$

which relates BLM modes and eigen modes of two tubules. To fit recent data [16] we used (11) and the hypothesis that the law $\omega_i \approx 228/d_i$ determines eigen frequencies of tubules in DWCNT (see Appendix B). The obtained fit shows the accuracy better than that of the model of coupled oscillators with two different masses used in [16] to fit these data. In principle, the model of coupled oscillators [16] can be rewritten in the form of Eq. (11). The difference between the fits appears due to the different assumptions about coupling between the tubes. We would also like to stress that our 2D continuous membrane approach is much more general one since, in contrast with the model [16], it describes coupling of all low-frequency modes in DWCNT, and not only the relation between BLM and a couple of RBM. Though, low-frequency modes in DWCNT are still not well studied experimentally, the proposed here model can constitute the basis for their further investigation.

Experimental study of the coupling constant $G'$ (which have the sense of coupling force per unit mass) and its dependence on diameters $d_1$ and $d_2$ of both tubules represents considerable interest for further identification and detailed characterization of DWCNT. Currently available experimental data [16] on BLM behavior is still limited and the dispersion of frequency values even for tubules which are quite similar (see, for example tubes 11 and 12, Table 1 in [16]) is rather high. In this situation it is worth to start the investigation of the tube-tube coupling with the approximation of the $G'(d_1,d_2)$ dependence by the simplest linear function. Fitting the data (see Appendix B) we obtain a function which gives standard deviation of 1.37 cm$^{-1}$ between experimental data and their fit, with the maximal deviation of 3.3 cm$^{-1}$. Further refinement of the fitting procedure leads to a quadratic function of $d_1$ and $d_2$ with the standard deviation of 1.31 cm$^{-1}$. In [16] "the unit-length coupling force constant owing to tube–tube interactions" with a complex dependence on $d_1$ and $d_2$ was introduced to characterize tubes coupling. Though this idea looks reasonable, it leads to a more complex fitting function and to a poorer fit with the standard deviation of 1.4 cm$^{-1}$ (see Appendix B).

The theory that we propose fits also with a good accuracy another recent Raman spectroscopy data [15] on DWCNT. In this work also, the behavior of coupled BLM was compared with the frequency variations of RBM in two co-axial SWCNT. The authors fitted the BLM behavior in the framework of a discrete model [18], and supposing that eigen frequencies of two tubules follow different laws as functions of their diameters, namely $\omega_{RBM1} \approx 204/d + 27$ for the frequency of the outer tubule, and $\omega_{RBM2} \approx 228/d$ for the frequency of the inner one. They explain this difference by the presence of impurity (amorphous carbon) on the outer tubule. In the fit performed in our work we suppose that impurity is present on both tubules and use the same law $\omega_{RBMi} \approx 204/d + 27$ for eigen frequencies of both tubules. Then, in the frame of the 2D continuous model the constant $G'$ is estimated as $G'$=1930 cm$^{-2}$nm and equation (11) gives for the frequencies of BLM in the DWCNT studied in [15] $\omega_1$ =185.6 cm$^{-1}$ and $\omega_2$ =133.4 cm$^{-1}$. The difference between calculated values and experimentally measured frequencies is less than 0.5 cm$^{-1}$, which is nearly one order of magnitude better than for the values obtained in the frame of the discrete model [18].

For both series of experimental Raman studies of DWCNT [15, 16] the measured frequencies of the couple of BLM in DWCNT differ only slightly from the RBM eigen frequencies of corresponding co-axial SWCNT. On the other hand, in the frame of our theory the estimated value of the *G'* constant is one order of magnitude smaller than the values $R_1\omega_1^2$ or $R_2\omega_2^2$. Both numerical values of frequencies and analysis of Eq. (11) for the estimated values of constants show then that the main contribution to the first mode in the couple of BLM is given by the vibrations of the outer wall of DWCNT, and the main contribution to the second mode corresponds to the inner wall vibrations. This conclusion illustrates BLM dynamics in DWCNT much better than rather popular idea that BLM can be classified as sin-phase and counter-phase oscillations of coupled co-axial tubules. The latter idea would have more sense in a couple of identical oscillators and not in a system of coupled tubules of different diameters.

In conclusion let us stress that the results obtained in the present work open the possibility for more detailed or alternative methods of SWCNT identification and characterization based on the frequency dependence on diameter for low-frequency Raman-active but non-resonant vibrational modes. The proposed theory applied to the case of DWCNT fits with a very good accuracy recent experimental data on BLM dynamics in DWCNT. The theory can be easily generalized to the case of triple- or multiple-walled carbon nanotubes (or other low-dimensional nanostructures with non-trivial topology) and applied to quantitative studies of their low-frequency vibrational spectrum, heat capacity and heat transfer properties.

Authors are grateful to J.-L. Sauvajol for fruitful discussions. V.L. acknowledges financial support of the Laboratory of Excellence NUMEV. S.B. acknowledges financial support of the RFBR grant 13-02-12085.

**Appendix A: 2D strain tensor**

The deformation of a 2D cylindrical shell in a 3D space is characterized by its 2D strain tensor. According to the principles of continuous media mechanics, such a deformation is dependent on a 3D local displacement field $\mathbf{u}=(u_r, u_\varphi, u_l)$ parameterized in cylindrical coordinates by the angle φ and the variable *l*, which measures the distance along the axis of the unstrained membrane. This displacement field is expressed in Cartesian coordinates as

$u_x^C = u_r \cos\varphi - u_\varphi \sin\varphi$

$u_y^C = u_r \sin\varphi + u_\varphi \cos\varphi$

$u_z^C = u_l$ ,

Equation $\mathbf{R'} = \mathbf{R}^0 + \mathbf{u}$ relates the initial Cartesian coordinates ($R\cos\varphi, R\sin\varphi, l$) of a material point $\mathbf{R}^0$ on the equilibrium cylindrical surface of radius *R* with its final coordinates $\mathbf{R'}$ on the deformed surface. The final positions of all points given by the dependence $\mathbf{R'}(\varphi, l)$ determine the shape of the deformed 2D shell. Then, the simplest way to introduce the strain tensor is to express it in terms of the metric tensor of the surface. On the one hand, following [8], the squared distance $dL'^2$ between two infinitely close points on the deformed shell surface is expressed as $dL'^2 = dL_0^2 + \varepsilon_{ij}\delta L_i \delta L_j$, where $dL_0^2 = \delta_{ij}\delta L_i \delta L_j$ is the squared distance between the same points in the initial unstrained state, $\varepsilon_{ij}$ is the 2D strain tensor, $\delta_{ij}$ is the Kronecker symbol, $\delta L_1 = \delta L_\varphi = R d\varphi$ and $\delta L_2 = \delta L_l$ are the distances along the cylinder transverse circular section perimeter and cylinder axis, respectively.

On the other hand, the squared distance $dL'^2$ defines a metric tensor [7]: $dL'^2 = g_{ij} d\upsilon_i d\upsilon_j$, where $d\upsilon_1 = d\varphi$ and $d\upsilon_2 = dL_l$, tensor $\widehat{g}$ being dependent on the displacement field. Finally, using deviation of the normalized metric tensor, we obtain the 2D local strain tensor $\varepsilon_{ij} = \frac{1}{2}(\gamma_{ik} g_{kl} \gamma_{lj} - \delta_{ij})$, where $\widehat{\gamma} = \widehat{g}_0^{-1/2}$ is the normalization, and $\widehat{g}_0$ is the metric tensor of the unstrained cylindrical membrane. In the explicit form $g_{12}= g_{21}=0$, $g_{11}=1/R$ and $g_{22}=1$. In the linear dynamics approximation studied in the present work, the strain is considered to be small. Thus, the nonlinear terms of the 2D strain tensor are neglected and its linear part takes the following form:

$$\varepsilon_{ij} = \begin{bmatrix} \partial_\varphi u_\varphi / R + u_r / R & \frac{1}{2}(\partial_l u_\varphi + \partial_\varphi u_z / R) \\ \frac{1}{2}(\partial_l u_\varphi + \partial_\varphi u_z / R) & \partial_l u_z \end{bmatrix}.$$

**Appendix B: Fit of BLM frequencies in DWCNT**

Table 1. BLM frequencies. Experimental data and values calculated in the frame of the proposed approach (columns $\omega_L^{theor}$ and $\omega_H^{theor}$) and using the model of Ref. 16 (two last columns).

| Sample | $d_1$ | $d_2$ | $\omega_L^{exp}$ | $\omega_H^{exp}$ | $\omega_L^{theor}$ | $\omega_H^{theor}$ | $\omega_L^{theor}$ [16] | $\omega_H^{theor}$ [16] |
|---|---|---|---|---|---|---|---|---|
| 1 | 3.17 | 2.46 | 79 | 115 | 78.9 | 114.8 | 78.7 | 113.0 |
| 2 | 2.60 | 1.90 | 98 | 137 | 95.9 | 137.2 | 95.9 | 137.2 |
| 3 | 2.33 | 1.64 | 108 | 155 | 106.7 | 154.9 | 106.9 | 155.4 |
| 4 | 3.03 | 2.30 | 83 | 112 | 81.6 | 113.0 | 81.5 | 112.6 |
| 5 | 2.94 | 2.20 | 84 | 114 | 83.1 | 113.6 | 83.2 | 113.9 |
| 6 | 2.85 | 2.12 | 85 | 120 | 86.3 | 119.5 | 86.4 | 119.8 |
| 7 | 2.64 | 1.92 | 93 | 129 | 93.3 | 131.2 | 93.5 | 131.8 |
| 8 | 2.31 | 1.62 | 110 | 156 | 107.6 | 156.4 | 107.7 | 156.9 |
| 9 | 2.71 | 2.01 | 93 | 133 | 92.2 | 132.1 | 92.2 | 131.8 |
| 10 | 2.47 | 1.76 | 103 | 143 | 99.9 | 142.6 | 100.1 | 143.3 |
| 11 | 2.38 | 1.71 | 103 | 154 | 105.6 | 154.6 | 105.6 | 154.5 |
| 12 | 2.38 | 1.71 | 105 | 155 | 105.6 | 154.6 | 105.6 | 154.5 |
| 13 | 2.33 | 1.64 | 108 | 154 | 106.7 | 154.9 | 106.0 | 155.4 |

We start the investigation of the tube-tube coupling with the approximation of the $G'(d_1,d_2)$ dependence by the simplest linear function. Fitting the data we obtain $G'=29700-43000d_1+44900d_2$, where $d_1$ is the outer wall diameter and $d_2$ is the inner wall diameter. This function gives standard deviation of 1.37 cm$^{-1}$ between experimental data and their fit, with the maximal deviation of 3.3 cm$^{-1}$. It is useful to compare this fit with that based on the coupling characteristics [16] (see Table 1). The authors of this work introduce "the unit-length coupling force constant owing to tube–tube interactions" directly proportional to the product of the mean diameter of the DWCNT and a linear function of distance between the walls. In the frame of the proposed approach it results in a more complex fitting function $G'(d_1,d_2)=[A+B(d_1-d_2)](d_1+d_2)$, where $A=6660$ and $B=-8480$. Though the idea [16] looks reasonable and interesting for applications, it leads to a more complicated fitting function and to a poorer fit with the standard deviation of 1.4 cm$^{-1}$. Further refinement of fitting procedure leads to a quadratic function $G'(d_1,d_2)=[A+B(d_1-d_2)+C(d_1+d_2)](d_1+d_2)$, where $A=7210$, $B=-9670$ and $C=61$, with the standard deviation of 1.31 cm$^{-1}$. Corresponding frequencies are presented in columns $\omega_L^{theor}$ and $\omega_H^{theor}$. Note that the same experimental data are fitted with a rather good accuracy by the quite different functions $G'(d_1,d_2)$. In our opinion this fact will stimulate further experimental studies of DWCNT with more pronounced difference in values of $d_1$ and $d_2$ parameters.